\documentclass[%
reprint,
superscriptaddress,
frontmatterverbose, 
 amsmath,amssymb,
 aps,
pra,
]{revtex4-2}
\usepackage{amsmath}
\usepackage{amssymb}
\usepackage{graphicx}
\usepackage{dcolumn}

\usepackage{bm}
\usepackage{hyperref}
\usepackage[mathlines]{lineno}
\usepackage{color}

\newcommand{\edit}[1]{{\color{black}#1}}
\newcommand{\newedit}[1]{{\color{black}#1}}

\begin{document}

\preprint{Synchronization and self-assembly of free capillary spinners
}

\title{Synchronization and self-assembly of free capillary spinners
}

\author{Nilgun Sungar}
 \email{nsungar@calpoly.edu}
\affiliation{
Department of Physics,
California Polytechnic State University, San Luis Obispo, San Luis Obispo, CA 93407}

\author{John Sharpe}
\affiliation{
Department of Physics,
California Polytechnic State University, San Luis Obispo, San Luis Obispo, CA 93407}

\author{Loic Ijzerman}
\affiliation{
Department of Physics,
California Polytechnic State University, San Luis Obispo, San Luis Obispo, CA 93407}

\author{Jack-William Barotta}
 \email{jack-william\_barotta@brown.edu}
\affiliation{
School of Engineering, Center for Fluid Mechanics,
Brown University, 184 Hope Street, Providence, RI 02912}

\date{\today}
\maketitle

\section*{Abstract}

\textbf{Chiral active particles are able to draw energy from the environment to self-propel in the form of rotation. We describe an experimental arrangement wherein chiral objects, spinners, floating on the surface of a vibrated fluid rotate due to emitted capillary waves. We observe that pairs of spinners can assemble at quantized distances via the mutually generated wavefield, phase synchronize and, in some circumstances, globally rotate about a point midway between them. A mathematical model based on wave-mediated interactions captures the salient features of the assembly and synchronization while a qualitative argument is able to rationalize global rotations based on interference and radiation stress associated with the wavefield. Extensions to larger collections are demonstrated, highlighting the potential for this tabletop system to be used as an experimental system capable of synchronizing and swarming.}

\section*{Introduction}

Self organization of interacting chiral active particles in space and time  has received much interest over the last two decades, with several review \cite{bishop2023active, shankar2022topological, lowen2016chirality} and perspective \cite{bowick2022symmetry, liebchen2022chiral} articles available. In these systems, particles expend energy which is either endogenously generated \cite{tan2022odd} or drawn from the environment \cite{grzybowski2002dynamic,workamp2018symmetry, bililign2022motile, soni2019odd, mcneill2023three} to move and rotate. Coupling between the particles \textemdash\ which leads to behavior such as collective rotation \textemdash\ can take several forms. For example, in the case of magnetically driven particles \cite{massana2021arrested, soni2019odd, bililign2022motile, chen2024self} the coupling is mediated by hydrodynamic forces while in the case of air driven spinners \cite{workamp2018symmetry} it is momentum transfer through steric interactions. \edit{Besides elucidation of the mechanisms driving and organizing these systems, study of the interaction and collective behavior of externally driven elementary particles may also find application in micro-robotics and targeted delivery \cite{wang2018collective, wang2021external}}

In this paper we present an experimental study of the self-organization and synchronization of untethered chiral particles at the air-liquid interface wherein interactions are wave-mediated.

An important example of wave-mediated interaction at the millimetric scale is liquid droplets bouncing at the air-liquid interface of a vibrated bath of fluid \cite{couder2005walking,bush2015pilot, bush2020hydrodynamic}. This system has yielded a plethora of phenomena reminiscent of those of quantum particles including diffraction\cite{couder2006single, pucci2018walking}, particle-in-a-box-like confinement \cite{harris2013wavelike,saenz2018statistical}, optical trapping \cite{sungar2017hydrodynamic}, and Anderson localization \cite{abraham2024anderson}. Emergent behavior has also been seen wherein droplets carried along by surface waves, and interacting through those waves, organize into hydrodynamic spin-lattices capable of synchronizing droplet-droplet phases and exhibiting both ferromagnetic and anti-ferromagnetic order \cite{saenz2021emergent}.

Recently, the addition of \textit{solid} bodies onto the vibrating liquid interface has allowed for control over particle shape to factor into wave-mediated self-propulsion and interactions \cite{ho2023capillary, oza2023theoretical, rhee2022surferbot, barotta2023bidirectional, barotta2024synchronization}. This is the case for “surfers” which are small solid, rectangular particles floating on the liquid \cite{ho2023capillary}. The surfers have an asymmetric mass distribution so that one side sits lower in the liquid than the other. Upon vibrating the fluid, the inertia of the surfer causes emission of capillary waves while the mass asymmetry creates an asymmetry in the strength (amplitude) of these waves. The difference in wave amplitude provides an imbalance in momentum transfer and thus propels the particle. When several particles are placed on the fluid, they are found to interact with each other via the emitted surface waves. Theory has been developed which accounts for the observed interactions \cite{oza2023theoretical}.

Considerations of asymmetry in the particle design suggest that a further degree of freedom may be probed – that of chirality. As first demonstrated by \cite{barotta2023bidirectional}, geometrically asymmetric particles, referred to as spinners, can acquire angular momentum from the wavefield. Furthermore, through careful design of the spinner geometry, the direction of rotation can be reversed by simply changing the wavelength of the emitted waves via modulating the frequency of vibration of the bath. Semi-quantitative arguments based on the balance of spinner size to wavelength demonstrate that the spatial structure of  wave interference is sufficient to account for the bidirectionality. This work has been extended to consider the case when two or more spinners are near each other \cite{barotta2024synchronization}. In this previous study, to keep focus solely on rotational dynamics, spinners are held at a fixed distance via magnetic restoring forces, and it is found that identical spinners can synchronize their phase. A companion model is able to explain the dynamic and static modes which persist. However, the study did not consider the case of untethered spinners wherein translational dynamics may allow for both synchronization and self-organization to occur.

In  this paper, we extend previously reported results on capillary spinners with several notable advances. By using a design which incorporates rotational mass asymmetry, we first validate the scaling reported for geometrically asymmetric prior designs \cite{barotta2023bidirectional}. Moreover, by placing multiple, untethered spinners in the bath, we are able to recover synchronization phenomena as previously reported \cite{barotta2024synchronization}. Furthermore, untethered spinners self-organize at quantized distances from one another at integer and half-integer wavelength spacing. Such bonded pairs of spinners admit global rotations while maintaining their fixed distance and phase difference. A companion model of wave-mediated interactions is able to rationalize such behavior of assembly in space and time, and a qualitative argument based on transverse forces provide intuition on the global rotation observed in experiment. Collections of spinners beyond the pair are presented as a promising candidate to higher collections wherein the system both "syncs" and "swarms" as an experimental "Swarmalator" system.

\section*{Experimental Setup}

\begin{figure}[ht!]
    \centering
    \includegraphics{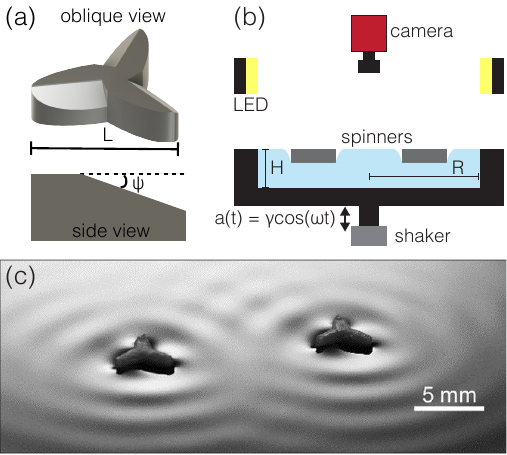}
    \caption{ \textbf{Overview of the experimental setup}. \textbf{(a)} Geometric design of the propeller-like spinner showing how the surface in contact with the liquid is cut at a slant $\psi$. The design is presented upside-down in order to highlight the geometry present on the bottom of the object. \edit{When rotating on the surface of a fluid, the slanted edge leads the rotation.} \textbf{(b)} The spinners are vertically vibrated at an acceleration of $a(t) = \gamma\cos(\omega t)$ in a water-glycerol bath of depth $H$ and radius $R$. A camera mounted above records the spinners with illumination provided by a ring of  LEDs $5$ cm above the fluid surface. \textbf{(c)} Oblique snapshot of two $L = 4.23$ mm spinners in the bath, slanted surfaces facing down. Outwardly propagating capillary waves generated by the spinners lead to both self-propulsion and interaction between neighboring spinners.
    }
    \label{fig: Setup}
\end{figure}

The spinners are 3-D printed on a resin printer (Formlabs Form2 with Grey Resin V4) and coated with a hydrophobic spray (Out Door Heavy Duty Protector, Adcor Products) before they are released into the bath. Once the bath begins to vertically oscillate at an angular frequency $\omega = 2\pi f$, the spinners both begin to emit outwardly propagating capillary waves and spontaneously start rotating around their center of mass in the direction determined by their chirality (Fig. \ref{fig: Setup}(a)),\edit{where the slanted edge of each wing leads the rotation.} The speed of rotation is constant, when the spinner is in isolation, and depends sensitively on fluid and driving parameters as well as the design of the spinner shape\cite{barotta2023bidirectional}. 

The experimental setup is shown in Fig. \ref{fig: Setup}(b). The bath of diameter $2R = 10.6$ cm and depth $H=5.7$ mm is made from laser-cut acrylic, with a water-glycerol (65-35 by volume, $\rho = 1099$ kg/m$^3$) mixture and the bath is fully filled to avoid the formation of a meniscus at the boundary. The bath is vertically shaken at an acceleration $a(t) = \gamma\cos(\omega t)$ using a Brüel and Kjær type 4809 vibration exciter powered by a type 2719 power amplifier. Driving signals are obtained from a GPIB-interfaced Agilent 33120A function generator. The bath's acceleration is monitored using two model 352C65 accelerometers attached to the bottom of the bath and a model 482C05 signal conditioner from PCB Piezotronics. Acceleration data is acquired through a National Instruments USB X multifunction DAQ with a control loop implemented via MATLAB ensuring the acceleration is within  $\pm 0.005$ g of the value set. A camera above the bath records videos at $20$ frames-per-second (fps).  The lighting is provided by a $24$ cm diameter ring of LEDs $5$ cm above and concentric with the bath, adjusted to produce a uniform dark background of fluid against the light gray spinners for tracking. The position and angular orientation of the spinners are obtained for each frame using an in-house program written in MATLAB.

\section*{The Single spinner}

The shape of the spinners used in our experiments is shown in Fig. \ref{fig: Setup}(a). The slanted bottom surface of one edge of each of the three wings creates a difference in the equilibrium deformation of the fluid surface around the wing. As the bath vibrates,  the relative motion of the spinner to the bath generates outwardly propagating capillary waves from the edges of the wings. The radiation stress generated by the waves thus shows an asymmetry between opposite edges of the wings creating a net force perpendicular to each wing \cite{roh2019honeybees,ho2023capillary, barotta2023bidirectional}. These non-radial forces generate a torque on the spinner causing it to rotate, with the slanted edge of the wings leading the rotation. The angle of slant, $\psi$, is a control parameter which can be adjusted to give a range of angular speeds for a fixed fluid acceleration. 

\begin{figure}
    \centering
    \includegraphics{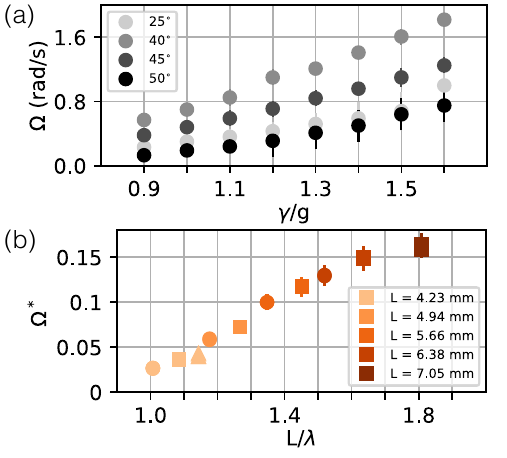}
    \caption{\textbf{Motion of a single spinner in isolation.} 
    \textbf{(a)} Resultant angular velocity as a function of the driving acceleration is measured for multiple slant angles. The maximal angular velocity is obtained for a slant angle of $40^{\circ}$ at a driving frequency of $f=70$ Hz and spinner size $L=4.23$ mm. \textbf{(b)} The dimensionless angular velocity is plotted as a function of the dimensionless size of the spinner $L/\lambda$ verifying the balance between wave and viscous stresses with circles ($\circ$), squares ($\square$), and triangles ($\triangle$) denoting $70$ Hz, $80$ Hz, and $90$ Hz, respectively \cite{barotta2023bidirectional}. See Supplemental Material (SM) for raw data used in verifying the collapse \cite{supp}. Error bars denote 2 standard deviations in the measured quantity.}
    \label{fig: SingleSpinner}
\end{figure}

At fixed frequency of $f=70$ Hz, the resultant angular velocity of the spinner, $\Omega$, is measured as a function of the amplitude of the driving acceleration $\gamma$ for a range of slant angles (Fig. \ref{fig: SingleSpinner}(a)). Rotation speed varies nonmontonically with a slant angle of the wings of $40^{\circ}$ generating the largest angular velocities for all of the accelerations tested. 

Such non-monotonic trends are to be expected when using chiral designs for wave-propelled objects \cite{barotta2023bidirectional}. In prior investigations on spinners in isolation, it has been shown that driving, fluid, and geometric parameters all contribute to the observed angular rotation speed. In the prior study, geometric asymmetry was used in the form of a chiral design to induce rotation in contrast to the design shown here which makes use of mass asymmetry on an achiral design. Nonetheless, we demonstrate that the rotation speed can still be rationalized with the same argument put forth in \cite{barotta2023bidirectional} by considering the balance between the driving torque arising from the wave-induced forces, and hence torques, on the wings and the resistive torque from a linear viscous drag. We define a dimensionless angular speed:

\begin{equation}
    \Omega^* = \frac{\Omega}{\left[ \frac{H\gamma^2\rho^2}{\mu\sigma k^4 L^2}\right]}
\end{equation}

where $\rho,\mu,\sigma$ are the density, dynamic viscosity  and surface tension of the fluid, respectively. The fluid depth is $H$ and the spinner size is $L$, the acceleration amplitude of the fluid is $\gamma$ and $k$ is the wavenumber of the capillary waves on the fluid surface. Given the relatively high frequencies used ($f > 60$ Hz), we may relate the wavenumber $k$ to the shaker angular frequency $(\omega = 2\pi f)$ of propagating capillary waves given by the deep-water dispersion relation, $\omega^2 = \sigma k^3/\rho$ \cite{lamb1924hydrodynamics}. This dimensionless angular speed should be proportional to a function $F(L/\lambda)$ which only depends on $L/\lambda$ for a fixed spinner shape where $\lambda$ is the wavelength of the capillary waves.

In Fig. \ref{fig: SingleSpinner}(b), we plot the experimental dimensionless angular speed for different values of dimensionless size $L/\lambda$ obtained by changing the size of the spinners ($L$), frequency (which in turn alters the wavelength), and the driving acceleration $(\gamma)$. The collapse of the data to a single curve confirms that the scaling arguments introduced in \cite{barotta2023bidirectional} also applies to our spinners (see Supplemental Material (SM) for raw data \cite{supp} \newedit{(including references \cite{barotta2023bidirectional})}. However contrary to the aforementioned study, the spinners presented here do not exhibit a flipping of the rotation direction  as the torque balance is always on the "heavier" side of the wings leading to $\Omega^*$ having a monotonic increase with $L/\lambda$.

With a characterization of the single spinner design established, we now move to interactions of free spinners coupled via the shared liquid interface.

\section*{Pair Interactions}

\begin{figure*}
    \centering
    \includegraphics{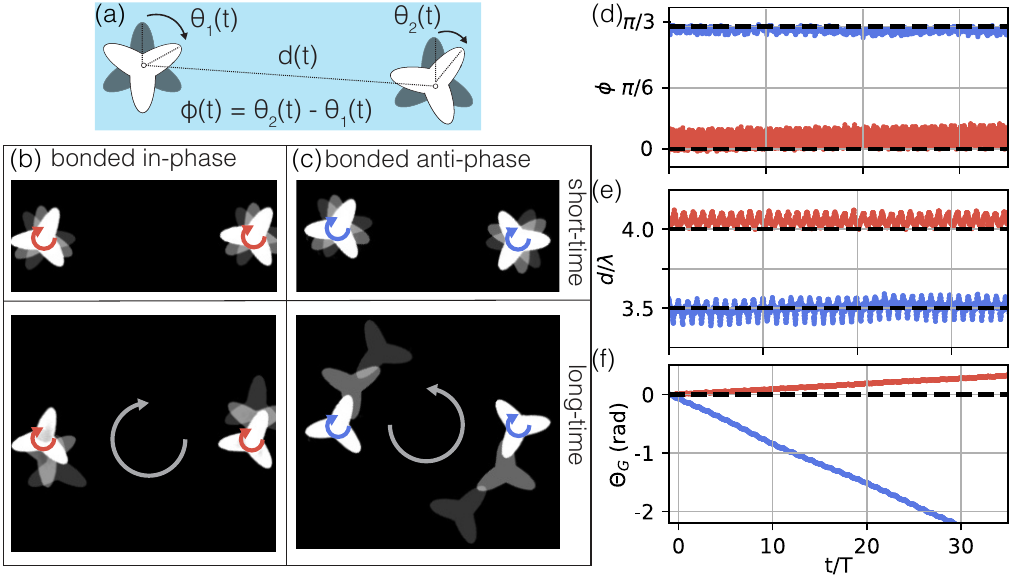}
    \caption{\textbf{Spinners organize at quantized distances and synchronize their rotations}. \textbf{(a)} A pair of spinners are defined by their positions and angle of rotation. A pair can then be expressed in terms of the center-to-center distance, $d(t)$, and the phase difference, $\phi(t) = \theta_2(t)-\theta_1(t)$. \textbf{(b)} A pair of clockwise spinners at $d/\lambda \approx 4$, rotating at $\Omega =1.03$ rad/s , exhibit a bonded in-phase mode wherein a phase difference of $\phi \approx 0$ is observed (top). Over longer periods of time, the bonded in-phase pair globally rotates about its center-point in the \textit{same} direction as the individual rotations  with angular speed $\Omega_G  =0.002$ rad/s (bottom). See Supplementary Video 1 for the corresponding movie. \textbf{(c)} A pair of clockwise spinners at a $d/\lambda \approx 3.5$, rotating at $\Omega =1.52$ rad/s exhibit a bonded anti-phase mode with a phase difference of $\phi \approx \pi/3$ (top). The bonded anti-phase pair globally rotates about its center-point in the \textit{opposite} direction as the individual rotations at $\Omega_G =0.02$  rad/s (bottom). See Supplementary Video 2 for the corresponding movie. Tracking the phase difference \textbf{(d)} effective distance, $d/\lambda$ \textbf{(e)}, and global rotation, $\Theta_G$ \textbf{(f)}, demonstrates stable, periodic, behavior over 30 full rotation periods, $T = 2\pi/\Omega$, for the spinners in both bonded pair modes. The global rotation has different directions for in-phase and anti-phase modes and the speed of the global rotation is higher in the anti-phase pair as can be seen from the slopes of the global angle graph \textbf{(f)}. }
    \label{fig: PairInteracts}
\end{figure*}

A pair of \textit{fixed} spinners are able to interact via their mutually generated wavefield by inducing torques on one another \cite{barotta2024synchronization}. To understand the interaction of an \textit{untethered} spinner pair, we track the position of the center of mass $(x_i,y_i)$ and orientation $(\theta_i)$ of each spinner at each video frame allowing for both the phase difference, $\phi(t) = \theta_2(t)-\theta_1(t)$ and separation distance $d(t)$ to be readily computed from this data (Fig. \ref{fig: PairInteracts}(a)). When two spinners of size $L 	\lessapprox  \lambda$   are placed in the vibrating bath the spinners are able to \edit{form a stable pair}, via adjusting their center-to-center distance, and synchronize their rotation, demonstrated via a fixed mean phase difference over time. This is dictated solely by the distance to wavelength ratio. \newedit{In the case of "spinners" without slants which do not rotate,  they do align their orientation to reach a phase difference of $\phi \approx \pi/3$ if they are close enough, but they do not show any interactions beyond that. } In the absence of waves the spinners in close proximity always collapse due to the Cheerios effect \cite{vella2005cheerios}. 

Stable bound pairs of spinners with the same or opposite chirality can be formed. This occurs when the spinner size $L$ is approximately equal to or smaller than the capillary wavelength, $\lambda$. We observe two distinct bound states: bonded and locked pairs. \edit{Bonded pairs occur when the mean separation distance remains constant while each spinner continues to rotate.  For locked pairs, the spinners cease to individually rotate while maintaining a fixed separation and orientation with respect to each other. In both bonded and locked pairs, global rotation is observed where the pair rotate about a point midway between them. }

We first discuss the bonded pair. For spinners of the same chirality, the pair also exhibits a global rotation (Fig. \ref{fig: PairInteracts}(b-c)) with the direction of global rotation dependent on the effective distance, $d/\lambda$, around a point midway between the spinners. The global rotation speed is much smaller than the individual spinner rotation speed $(\Omega_{\text{global}} \approx \Omega_{\text{individual}}/100 )$. Behavior is stable over long timescales with fluctuations about mean values of phase difference and effective distance, $d/\lambda$, resulting from wave-mediated torques (Fig. \ref{fig: PairInteracts}(d-e)).  The observed fluctuations in the speed are due to ever-changing edge-to-edge distance between the wings of the bonded pair. 

\begin{figure}
    \centering
    \includegraphics{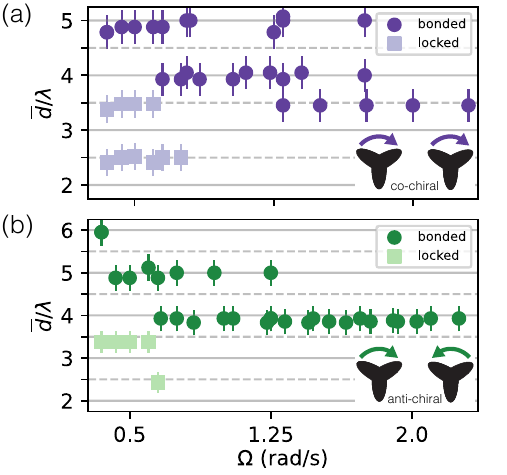}
    \caption{\textbf{Equilibrium mean distances are quantized by the wavelength}. (\textbf{a}) A pair of spinners of the same chirality equilibrate their mean center-to-center spacing, $\overline{d}$, to either be integer or integer and half wavelength spacing. When the effective distance, $\overline{d}/\lambda$ is small and the angular velocity is low, the locking bound state is achieved (light purple, square, markers). (\textbf{b}) A pair of spinners of opposite chirality also exhibit both bonded and locked pair states. Bonded pairs only appear at integer wavelength spacings. Data is for spinners of size $L = 4.23$ mm driven at $f=70$ Hz in a $65-35$ water-glycerol mixture (by volume). Different speeds are obtained by changing the acceleration and the slant angle. Error bars denote 2 standard deviations in the measured quantity.}
    \label{fig: DistPhaseResults}
\end{figure}

We can further separate distinct regimes within the broader class of bonded pairs. When the pair is placed in the bath, the spinners adjust their center-to-center distance to either integer multiples of the wavelength $(d\sim n\lambda)$ or integer and a half multiples of the wavelength $d\sim(n+1/2)\lambda$ (Fig. \ref{fig: DistPhaseResults}). The robustness of the bonding can be seen when we modulate the frequency in real time while a pair is in a bonded state. For example, with a bonded pair at 80 Hz separated by an integer number of wavelengths, we can rapidly change the frequency to 70 Hz. The pair quickly adjusts the separation to an integer multiple of the new  wavelength.  The data for this switch is shown in the  Supplemental Material (SM) \cite{supp}.  

For pairs with the same chirality, when the distance is approximately an integer multiple $(d\sim n\lambda)$, we find that the spinner pair finds an in-phase synchronization $(\phi \sim 0)$ (Fig. \ref{fig: PairInteracts}(d)(e)). Experimentally, bonded pairs have been observed for $n=4,5,6$ wherein distances are sufficiently large enough to still retain individual rotation i.e. the spinners do not lock. These bonded pairs display their slow global rotation around the midpoint between the spinners with the same direction as the spinner rotation (Fig. \ref{fig: PairInteracts}(b). 

Bonded pairs with the same chirality at separations $d \sim( n+1/2)\lambda$ where $n=3,4,5$ are also observed. Spinners again synchronize their rotation but with a phase difference of $\phi \approx \pi/3$ (Fig. \ref{fig: PairInteracts}(c)). The global rotation of these bonded pairs is opposite to the spinner rotation direction, in contrast to the global rotation of in-phase synchronized pairs.  The speed of the global rotation, which is only a few percent of the spinner rotation speed, is inversely  proportional to the separation of the pair and weakly proportional to the acceleration. This is expected as the interaction between the spinners will be stronger at smaller separations and for larger wave amplitudes.  

With pairs of opposite chirality, we only observed bonded pairs at separations that are integer multiples of the wavelength $(d \sim n\lambda)$ (Fig. \ref{fig: DistPhaseResults}(b)). For these bonded pairs, the center of mass of each spinner stays at rest and there is no global rotation.

The locked pairs are observed only when the angular speed of the spinners is below approximately $0.75$ rad/s and for separations of $d \sim(n+1/2) \lambda$ where $n=2,3$ (Fig. \ref{fig: DistPhaseResults}). This is a regime wherein the self-propulsion torque is relatively low, as indicated by the slow speed, and the wave interaction torques are relatively large because of the smaller spacing between the objects. For pairs with opposite chirality, the pair comes to rest upon locking and there is no global rotation. Both individual rotation and global rotation is zero. On the contrary, the locked pairs of spinners with the same chirality always have a global rotation in the opposite direction to the spinner rotation direction. This is akin to the behavior of bonded pairs assembled at half integer wavelength spacing. See Supplementary Video 3 for the corresponding movie of a locking event. 

\section*{A model for spinner interactions}

\begin{figure}
    \centering
    \includegraphics{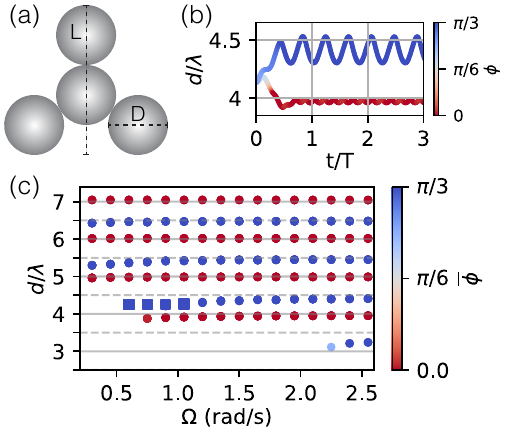}
    \caption{\textbf{Simulating pairwise interactions of free spinners}. \textbf{(a)} Each spinner is simulated as 4 point sources arranged to mimic the shape of the experimental spinner with $L=4.20$ mm and $D = 1.9$ mm. \textbf{(b)} A pair quickly adjusts their phase difference, \edit{$\phi = \theta_2-\theta_1$}, and pairwise distance, \edit{d}, to find an in-phase or anti-phase mode with integer or half integer wavelength spacing, respectively. \textbf{(c)} Simulating over a range of angular rotation rates, qualitative agreement is found between simulation and experiment wherein both bonded and locked modes are replicated with alternating in-phase and anti-phase modes of synchronization. \newedit{Circle markers denote bonded states whereas square markers denote locked states.}
    }
    \label{fig: SimOutput}
    \end{figure}

In order to rationalize the wave-mediated interactions, we simulate collections of untethered spinners. The governing equations of motion for both the force and torque have been previously derived both for untethered surfers \cite{oza2023theoretical} and tethered/fixed spinners \cite{barotta2024synchronization}. To summarize, we treat each spinner as a collection of point sources that can be defined by the lateral position $(x_n(t),y_n(t))$ and rotation angle $\theta_n(t)$. For simplicity, we choose to represent each spinner as $J=4$ point sources (Fig. \ref{fig: SimOutput}(a)); a point source is placed in the middle and a single point source is placed on each wing of the spinner. For each point source pair, both an attractive capillary force from the static deformation of the interface \cite{vella2005cheerios}, $F_{\text{c},ij}$ and dynamic oscillatory wave-mediated force is computed $F_{\text{w},ij}$ \cite{de2018capillary}. For the capillary attraction force, each point source is thought of as a disk of diameter $D = 4L/9$ ensuring that the approximate geometric description of the shape in simulation is comparable to the experiment.
\edit{ To model the interaction of two wave-emitting point particles, we prescribe the wave force on particle $i$ from the wavefield of particle $j$ by $\mathbf{F}_{ij}=\left\langle\left. m_i \ddot{z}_i \nabla h\left(\mathbf{x}-\mathbf{x}_j, t\right)\right|_{\mathbf{x}=\mathbf{x}_i}\right\rangle$ where $\langle \cdot \rangle$ represents the time average over one oscillation period. Each point source oscillates vertically and responds to the wavefield gradient, $\nabla h$, generated from the other point source \cite{ de2018capillary, oza2023theoretical}. The wave-mediated force depends on distance - decaying and oscillating over a period given by the wavelength. Additional details on the functional forms for the capillary attraction force, wave-mediated force, and wavefield expression are found in the supplemental material (SM) along with fluid and driving parameters used \cite{supp} \newedit{(including references \cite{de2018capillary, oza2023theoretical})}.}  The pairwise interactions are then summed over each of the spinners leading to a resultant force and torque given by the equations of motion 

\begin{equation}
 m\ddot{\mathbf{x}}_n +\frac{m}{{t}_{\nu}}\dot{\mathbf{x}}_n= \sum_{i=1}^{J}\sum_{j=1}^{J}  \left(\mathbf{F}_{\text{w},ij}+\mathbf{F}_{\text{c},ij}\right),
 \label{eq:odeF}
\end{equation}

\begin{equation}
 I\ddot{\theta}_n +\frac{I}{{t}_{\nu}}\dot{\theta}_n= \frac{I}{{t}_{\nu}}\Omega_n+ \sum_{i=1}^{J}\sum_{j=1}^{J} \left[ \mathbf{r}_{i} \times  \left(\mathbf{F}_{\text{w},ij}+\mathbf{F}_{\text{c},ij}\right)\right] \cdot \hat{\mathbf{z}},
 \label{eq:odeI}
\end{equation}

For a spinner pair initialized at an arbitrary distance, the pair quickly adjust their center-to-center distance to be near an integer or half integer wavelength spacing as seen in experiment with the phase difference corresponding to either the in-phase or anti-phase mode, respectively (Fig. \ref{fig: SimOutput}(b)). To compare directly to experiment, we considered a sweep of the resultant effective distance, $d/\lambda$, and mean phase difference, $\overline{\phi}$, as a function of the individual rotation of the spinner (Fig. \ref{fig: SimOutput}(c)). 

In qualitative agreement with experimental data (see Fig. \ref{fig: DistPhaseResults}), integer and half integer wavelength spacing is recovered over the entire range of spinner rotation rates. In addition, for relatively small effective distances, the locking mode is also observed (red squares). While the locking mode is observed at larger effective distances than experiment, the qualitative trends of  the effective distance and mean phase difference are all recovered indicating that the alternating behavior between modes and distances is a result of the wave-mediated interaction.

For low effective distances, we note that there is collapse of the spinner pair, similar to what is observed in experiment. This is directly due to the fact that the purely attractive force from the static menisci is effectively too large to balance out the oscillatory wave forces which are being exerted on the spinners. We have tested that when the static capillary attraction force is set to zero, collapse is avoidable, and the small effective distance and low angular velocity regime (bottom left of (Fig. \ref{fig: SimOutput}(c)) is filled with more locking modes at half integer wavelength spacing as expected.

While the simulation is able to recover the salient features of the experiment, the trends in the global rotation rate are \textit{not} recovered. For a pair of spinners of the same chirality, global rotations persist, but the direction of the global rotation is always in the same direction as the individual rotation. \edit{This is contrary to experimental observations where the global rotation direction depends on the spacing of the pair.} To rationalize the net global rotations, we turn to a qualitative argument based on wave interference.

\subsection*{Global rotations}
We infer from our observations that the interaction of spinners in a pair is mediated by capillary waves emitted by each spinner and the subsequent interference of these waves in the region between the pair. When a spinner is in the bath by itself, it spins at a uniform speed while its center stays at rest.  This indicates that the forces on the wings are equal, resulting in a net torque but zero net force.  

\begin{figure}
    \centering
    \includegraphics{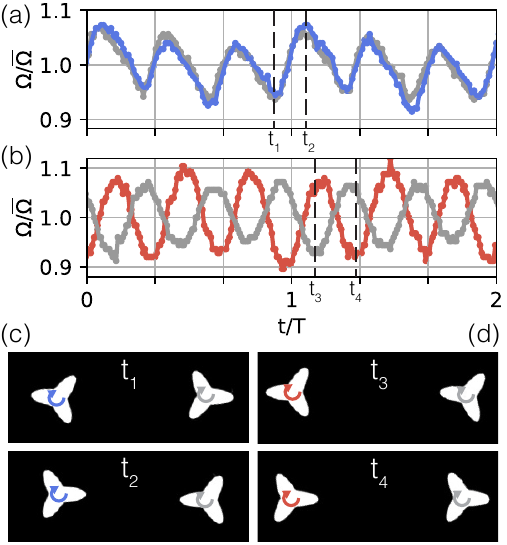}
    \caption{\textbf{Angular velocity response of a bonded pair} \textbf{(a)} The bonded anti-phase mode has fluctuations in angular velocity which are in tandem. The left spinner is shown in blue and right spinner is shown in gray. \textbf{(b)} The bonded in-phase mode has alternating fluctuations in angular velocity with the left spinner shown in red and the right spinner in gray. Fluctuations occur on a period of $T/3$ due to the three-fold symmetry of the object. Tracking the relative angular position of the spinner pair reveals that the position of the wedge in the inner region between spinners determines the fluctuations in speed for both anti-phase (\textbf{c}) and in-phase (\textbf{d}) synchronization.
    }
    \label{fig: Fig8}
\end{figure}

When a spinner is in a bonded pair, we find that its angular speed oscillates with a period that is $1/3$ of the spinning period.  Fig. \ref{fig: Fig8} shows the experimental angular speed for each spinner in a bonded pair for integer and half integer separations. For a pair separated by $d \sim n \lambda$,  oscillations of the angular speed of the spinners are out of phase with each other  (Fig. \ref{fig: Fig8}(b)). Inspection of the corresponding videos show that the angular speed decreases whenever a wing is moving into the region between the spinners. In this mode, due to the orientation of the spinners, the wings from each spinner alternately move through the region between the spinners (Fig. \ref{fig: Fig8}(d)). For a bonded pair separated by $d \sim (n+1/2) \lambda$, the oscillations of the spinners angular speed are in phase as shown in Fig. \ref{fig: Fig8}(a). In this mode the wings of the two spinners move through the region between the spinners at the same time (see Fig. \ref{fig: Fig8}(c)).
    
We can rationalize our experimental observations with a simple model of the forces and torques acting on each spinner in a bonded pair utilizing the interference of the waves in the region between the spinners.

We assume that the waves from the wing of one spinner will interfere constructively with the waves emitted from the other spinner when the spinners are integer wavelengths apart and interfere destructively when they are half-integer wavelengths apart.  This will change the radiation stress on the wing that is moving into the region between the spinners. This in turn changes the net torque on the spinners, changing the angular speed. The net force will now no longer be zero, causing the center of each spinner to translate.  The direction of the global rotation in bonded pairs can be explained by considering the direction of the force on each spinner for each mode.  

\begin{figure}
    \centering
    \includegraphics{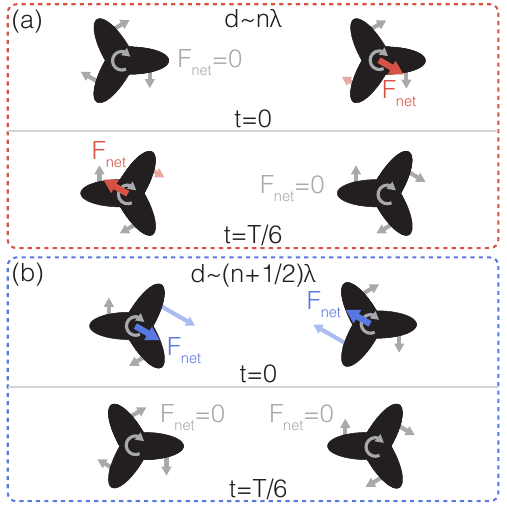}
    \caption{
    \textbf{Sketch of the forces acting on the wings of a bonded pair at two instances T/6 apart, where T is the spinning period.}  $ \textbf{(a)} $When the separation is $d \approx n\lambda$, the force on the wing approaching the central region (spinner on the right in top panel) decreases, causing a decrease in its rotation speed and a net force on that spinner in the direction shown by the red arrow. At this time the spinner on the left has no net force. T/6 later (bottom panel), the wing of the spinner on the left is approaching the middle region and the force on the wing has decreased, causing a decrease in its rotation speed and a net force in the  direction shown by the red arrow while the spinner on the right has no net force. $ \textbf{(b)}$When  $d \approx (n+1/2)\lambda$, wings of both spinners approach the central region at the same time (top panel), both experiencing an increase in the force, causing their angular speed to increase and a net force on each spinner in the directions shown by the blue arrows. At a time T/6 later (bottom panel), the forces on the wings will be equal for both spinners and there will be no net force on either one.} 
    \label{fig: Fig9}
\end{figure}

Fig. \ref{fig: Fig9} offers a schematic of two clockwise-rotating bonded spinners. In Fig. \ref{fig: Fig9}(a) the spinners are an integer wavelength apart and spinning in-phase, in Fig. \ref{fig: Fig9}(b) they are a half integer apart and spinning with a $\pi/3$ phase difference. \edit{The vectors sketched in the figure show the conjectured relative magnitudes of the forces on the wings and the direction of the net force. They do not represent the actual magnitudes of these forces, which are unknown.}

In the first case, constructive interference will increase the wave amplitude on the leading edge of the wing on the left-hand spinner which is moving into the region between the spinners. This will lead to a decrease in the \textit{net} radiation stress on the wing and decreases the force on that wing, causing a decrease in its angular speed, as observed. It will also generate a net force on the spinner which has an upward component for the spinner on the left. The right-hand spinner will experience a downward force as it's wing moves into the region, via an analogous argument. If we examine the whole cycle, the net force changes direction but maintains a component of force perpendicular to the line joining the spinners that would generate a global clockwise rotation.  This simple argument agrees with our observation that the slow global rotation is in the same direction as the spinner rotation for an in-phase pair. 

Fig. \ref{fig: Fig9}(b) sketches the forces on each spinner in a bonded pair separated by $d \sim(n+1/2) \lambda$ and rotating clockwise with a phase difference of $\pi/3$. In this case, destructive interference will decrease the wave amplitude at the leading edges of the wings which are both moving into the region between the spinners.  This will lead to an increase in the radiation stress on these wings, yielding an increase in the angular speed of both spinners. A net force is also generated that has a downward component for the spinner on the left and an upward component on the spinner on the right. Again, if we examine the whole cycle, the net force changes direction but maintains a component at right angles to the line joining the spinners. This agrees with our observation that the global rotation is in the direction opposite that of the spinner rotation. 

The simple pairwise interaction model of the wave-mediated forces used in the simulation does not capture such  phenomena due to the self-propulsion term being fixed rather than being a function of the phase difference. It is conjectured that by adapting the self-propulsive term to account for interference effects from neighboring spinners, the global rotation direction will be able to be captured. This is beyond the scope of the work presented here. We believe the wave-mediated forces modeled in simulation are likely the dominant effect at play, with the more subtle secondary effect of wave interference allowing for net rotation to arise. Indeed, such second order interference effects have been postulated to explain the behavior of millimetric chiral particles subject to airborne acoustic fields \cite{king2024nonreciprocal, st2023dynamics}.

\section*{Discussion and Outlook}

\begin{figure*}
      \centering
      \includegraphics{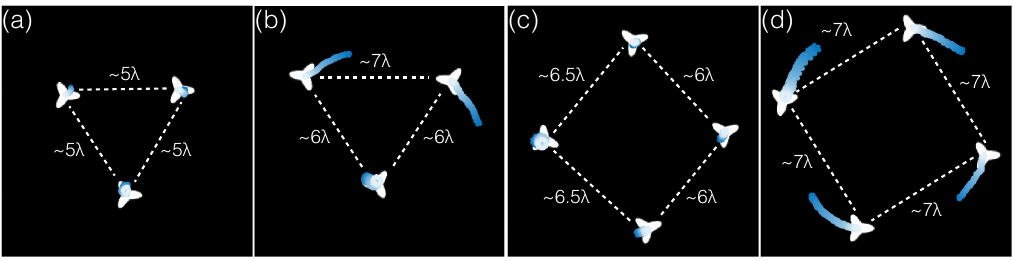}
      \caption{\textbf{Collective motion of free capillary spinners.} When $n=3$ spinners of the same chirality are placed in the bath, the dynamics of the group are dependent on the inter-particle spacing resulting in both globally static \textbf{(a)} and dynamic modes \textbf{(b)}. An analogous situation occurs for $n=4$ spinners \textbf{(c-d)}. Notably, when equally spaced at $d \sim 7\lambda$  from nearest neighbors, a square of four spinners demonstrates a collective global rotation.}
      \label{fig: NbodyModes}
  \end{figure*} 
We have presented an experimental system composed of untethered, capillary wave-driven spinners. We have demonstrated that the spinners can self-organize at integer and half-integer multiples of the capillary wavelength and synchronize the phase of their rotation. A further intriguing observation is that these “bonded” pairs of co-chiral spinners exhibit \textit{global} rotation about an axis midway between them where the global rotation direction is the same as the chirality of the spinners for integer wavelength separations and opposite to the chirality of the spinners for half-integer wavelength separations. For free capillary surfers \cite{ho2023capillary} stable bonded states with integer wavelength separations had also been observed. The new bonded states with half-integer wavelength separation observed in our system is likely the result of the varying edge-to-edge distance of the wings as the pair rotates, allowing for equilibrium mean separation distances that are not always full integer.  The surfer pairs interact through edges that are at rest relative to one another and the distance between the edges, hence pair separation does not change. The periodic fluctuations in  speed and separation observed for the spinner pairs, which is a feature not realized with the surfers, are due to periodically changing edge-to-edge distance between the wings.  

While the work presented here focuses on bonded pairs, “locked” pairs are also observed where the spinners halt their individual rotation while maintaining their global rotation. These states are only seen at low speeds and small, half-integer wavelength separations with the global rotation direction opposite to the chirality of the spinners.

An agent-based model with fluid-mediated interactions derived from first principles captures many of the salient features, such as self-organization and phase synchronization, but notably does not show bidirectional global rotation dependent on the interparticle spacing. We believe this disagreement is because the model does not adequately incorporate interference between the waves emanating from the spinners; an effect which would be captured with full radiation stress arguments using the total resultant wavefield of the collection of spinners. To remedy this shortcoming of the model, we have developed a qualitative explanation for the direction of the global rotation which aligns with experimental observations. This simple picture also correctly predicts the observed relative rotational velocity fluctuations of the individual spinners.

\newedit{Future investigations will seek to  understand the global rotations present due to spinner-spinner interactions and better understand why the direction of global rotation appears to depend sensitively on the effective distance between the spinners. It has been shown \cite{barotta2023bidirectional} that when the amplitude of the waves is small $(kA \ll 1$ where $A$ is the amplitude of the waves and $k$ is the wavenumber), waves are the dominant mechanism both for self-propulsion and interaction as evidenced throughout this study. However, when considering the slow global rotations present, it may be necessary to consider also the associated surface flows present on the surface \cite{roh2019honeybees, punzmann2014generation} and/or wave interferences outlined. We conjecture that one of these nonlinear mechanisms may be a driving factor in rationalizing such observations which occur on much longer timescales compared to the individual rotation rate. We anticipate that with the observations laid forth here, modeling efforts can subsequently follow in future investigations that may incorporate non-reciprocal forces from such wave interference and/or flow-mediated interactions which has been recently successful in rationalizing similar emergent activity in acoustic wave setups \cite{king2024nonreciprocal, st2023dynamics}.}

Recently, systems composed of interacting particles that can both organize in space and synchronize in time (“swarmalators”) have received much attention \cite{o2017oscillators}. While such systems can be created \textit{in silico} and may be observable in nature \cite{yuan2014gait, peshkov2022synchronized}, the experiments reported here have many advantages when it comes to table-top implementation and control of parameters: the chiral particles are readily manufactured by 3D printing and at a size congruent with the spatial dimensions of the capillary waves, permitting self-organization. The individual particles can be selected for chirality and, by variation of the slant angle, tuned to different rotation speeds allowing for the study of poly-disperse mixtures. \edit{The system presented herein is a candidate for testing the extensive theoretical predictions put forth within the swarmalator framework. With the possibility of modes such as static synchronization and a variety of phase waves (static, splintered, active) \cite{o2017oscillators,o2022collective}, we envision that by mapping the current system into the formalism of the swarmalators, experimental design can be informed by the currently standing theoretical predictions. In addition to the work functioning as a tunable experimental platform that can fit within the general swarmalator framework, it can also be used to test and help elucidate more general questions in chiral active matter. \cite{liebchen2022chiral,yang2020robust, caprini2024self, workamp2018symmetry}. To best probe the collective dynamics of chiral active matter, minimal models are necessary, wherein suitable analysis can be conducted. We hope that a connection between the swarmalator framework and the active matter community can be realized in this work.
 }

As an outlook to future studies moving beyond a pair and highlighting the complex interaction landscape, we present experimentally observed N-body spinner interactions between identical particles. (Fig. \ref{fig: NbodyModes}).

In Fig. \ref{fig: NbodyModes}(a), three equidistant spinners with the same chirality synchronize but do not display any global rotation. However, in the case where one spinner is displaced, creating an isosceles triangle, global rotation of the two spinners around the one at the apex is observed (Fig. \ref{fig: NbodyModes}(b)). Moving to four spinners, bound states can likewise be formed, exhibiting global rotation dependent on the separation and relative phases of the spinners (Fig. \ref{fig: NbodyModes}(c-d)). In both cases, individual rotation is retained while the global rotation of the spinner network depends sensitively on the inter-spinner distances. See Supplementary Video 4 for the corresponding movie of 4 globally rotating spinners.

In summary, the wave-mediated interaction coupled with the chiral design leads to a rich array of collective behavior that depends sensitively on interparticle distance and rotation rate of the spinners. We anticipate that the ability for such objects to both "sync" and "swarm" will allow for future analysis to fall within the general swarmalator framework,  allowing for fundamental questions in chiral active matter to be studied \cite{liebchen2022chiral}.
  
\section*{Acknowledgements}
The authors thank Daniel M. Harris and Giuseppe Pucci for stimulating discussions and useful feedback on the manuscript. N.S., J.S. and L.I. thank the Bill and Linda Frost Fund for support of this work. J.-W.B. is supported by the Department of Defense through the National Defense Science and Engineering Graduate (NDSEG) Fellowship Program and Brown University's Presidential Fellowship.

\section*{Author contributions}

N.S., J.S, and J.-W.B. designed research;
N.S., J.S., and L.I. conducted experiments; J.-W.B. developed and performed simulations; J.-W.B. and N.S. analyzed data and developed models. N.S., J.S., and J.-W.B. wrote the paper; all authors performed research, discussed the results, commented on the manuscript and gave final approval for publication, agreeing to each be held accountable for the work performed therein.

\section*{Data availability}

Experimental data in the study and the CAD files for the spinners are available from the corresponding author N.S. by reasonable request.

\section*{Code availability}

Simulation code is available from the corresponding author J.-W.B. by reasonable request.

\section*{Competing interests}
The authors declare no competing interests.

\section*{Additional information}
\textbf{Supplementary information} 4 Supplementary Videos and Supplementary Material (SM) are available.

\bibliography{refs}

\end{document}